\PassOptionsToPackage{unicode,colorlinks=true,linkcolor=blue,citecolor=blue,urlcolor=blue}{hyperref}
\PassOptionsToPackage{hyphens}{url}
\PassOptionsToPackage{dvipsnames,svgnames,x11names}{xcolor}
\documentclass[
  11pt,
  letterpaper,
  DIV=11,
  numbers=noendperiod]{scrartcl}

\usepackage{amsmath,amssymb}
\usepackage{setspace}
\usepackage{iftex}
\ifPDFTeX
  \usepackage[T1]{fontenc}
  \usepackage[utf8]{inputenc}
  \usepackage{textcomp} 
\else 
  \usepackage{unicode-math}
  \defaultfontfeatures{Scale=MatchLowercase}
  \defaultfontfeatures[\rmfamily]{Ligatures=TeX,Scale=1}
\fi
\usepackage{lmodern}
\ifPDFTeX\else  
\fi
\IfFileExists{upquote.sty}{\usepackage{upquote}}{}
\IfFileExists{microtype.sty}{
  \usepackage[]{microtype}
  \UseMicrotypeSet[protrusion]{basicmath} 
}{}
\makeatletter
\@ifundefined{KOMAClassName}{
  \IfFileExists{parskip.sty}{
    \usepackage{parskip}
  }{
    \setlength{\parindent}{0pt}
    \setlength{\parskip}{6pt plus 2pt minus 1pt}}
}{
  \KOMAoptions{parskip=half}}
\makeatother
\usepackage{xcolor}
\usepackage[margin=1in]{geometry}
\ifx\paragraph\undefined\else
  \let\oldparagraph\paragraph
  \renewcommand{\paragraph}[1]{\oldparagraph{#1}\mbox{}}
\fi
\ifx\subparagraph\undefined\else
  \let\oldsubparagraph\subparagraph
  \renewcommand{\subparagraph}[1]{\oldsubparagraph{#1}\mbox{}}
\fi

\providecommand{\tightlist}{
  \setlength{\itemsep}{0pt}\setlength{\parskip}{0pt}}\usepackage{longtable,booktabs,array}
\usepackage{calc} 
\usepackage{etoolbox}
\makeatletter
\patchcmd\longtable{\par}{\if@noskipsec\mbox{}\fi\par}{}{}
\makeatother
\IfFileExists{footnotehyper.sty}{\usepackage{footnotehyper}}{\usepackage{footnote}}
\makesavenoteenv{longtable}
\usepackage{graphicx}
\makeatletter
\def\maxwidth{\ifdim\Gin@nat@width>\linewidth\linewidth\else\Gin@nat@width\fi}
\def\maxheight{\ifdim\Gin@nat@height>\textheight\textheight\else\Gin@nat@height\fi}
\makeatother
\setkeys{Gin}{width=\maxwidth,height=\maxheight,keepaspectratio}
\makeatletter
\def\fps@figure{htbp}
\makeatother

\usepackage{booktabs}
\usepackage{longtable}
\usepackage{array}
\usepackage{multirow}
\usepackage{wrapfig}
\usepackage{float}
\usepackage{colortbl}
\usepackage{pdflscape}
\usepackage{tabu}
\usepackage{threeparttable}
\usepackage{threeparttablex}
\usepackage[normalem]{ulem}
\usepackage{makecell}
\usepackage{xcolor}
\usepackage{amsbsy}
\usepackage{amsmath}
\usepackage{amssymb}
\usepackage{amsthm}
\usepackage{booktabs}
\usepackage{threeparttable}
\usepackage{multirow}
\usepackage{dcolumn}
\newtheorem{assumption}{Assumption}
\newtheorem{proposition}{Proposition}
\newtheorem{remark}{Remark}

\KOMAoption{captions}{tableheading}
\makeatletter
\@ifpackageloaded{caption}{}{\usepackage{caption}}
\AtBeginDocument{
\ifdefined\contentsname
  \renewcommand*\contentsname{Table of contents}
\else
  \newcommand\contentsname{Table of contents}
\fi
\ifdefined\listfigurename
  \renewcommand*\listfigurename{List of Figures}
\else
  \newcommand\listfigurename{List of Figures}
\fi
\ifdefined\listtablename
  \renewcommand*\listtablename{List of Tables}
\else
  \newcommand\listtablename{List of Tables}
\fi
\ifdefined\figurename
  \renewcommand*\figurename{Figure}
\else
  \newcommand\figurename{Figure}
\fi
\ifdefined\tablename
  \renewcommand*\tablename{Table}
\else
  \newcommand\tablename{Table}
\fi
}
\@ifpackageloaded{float}{}{\usepackage{float}}
\floatstyle{ruled}
\@ifundefined{c@chapter}{\newfloat{codelisting}{h}{lop}}{\newfloat{codelisting}{h}{lop}[chapter]}
\floatname{codelisting}{Listing}

\makeatother
\makeatletter
\@ifpackageloaded{caption}{}{\usepackage{caption}}
\@ifpackageloaded{subcaption}{}{\usepackage{subcaption}}
\makeatother
\ifLuaTeX
  \usepackage{selnolig}  
\fi
\usepackage[]{natbib}
\usepackage{bookmark}

\IfFileExists{xurl.sty}{\usepackage{xurl}}{} 
\hypersetup{
  pdftitle={Bandwidth Selection for Spatial HAC Standard Errors},
  pdfauthor={Alexander Lehner},
  colorlinks=true,
  linkcolor={blue},
  filecolor={Maroon},
  citecolor={Blue},
  urlcolor={Blue},
  pdfcreator={LaTeX via pandoc}}

\title{Bandwidth Selection for Spatial HAC Standard Errors}
\author{Alexander
Lehner\thanks{World Bank Group, Outcomes Department, 1818 H Street, Washington, DC. Email: alehner@worldbank.org. I thank audiences at UChicago for helpful comments and discussions. Alex Gordon provided excellent research assistance. Computations were carried out on the Akropolis cluster at the University of Chicago. The findings, interpretations, and conclusions expressed in this paper are entirely those of the author and do not necessarily represent the views of the World Bank Group, its Executive Directors, or the countries they represent.}}
\date{2026-03-03}

\begin{document}
\maketitle
\begin{abstract}
Spatial autocorrelation in regression models can lead to downward biased
standard errors and thus incorrect inference. The most common correction
in applied economics is the spatial heteroskedasticity and
autocorrelation consistent (HAC) standard error estimator introduced by
Conley (1999). A critical input is the kernel bandwidth: the distance
within which residuals are allowed to be correlated. However, this is
still an unresolved problem and there is no formal guidance in the
literature. In this paper, I first document that the relationship
between the bandwidth and the magnitude of spatial HAC standard errors
is inverse-U shaped. This implies that both too narrow and too wide
bandwidths lead to underestimated standard errors, contradicting the
conventional wisdom that wider bandwidths yield more conservative
inference. I then propose a simple, non-parametric, data-driven
bandwidth selector based on the empirical covariogram of regression
residuals. In extensive Monte Carlo experiments calibrated to
empirically relevant spatial correlation structures across the
contiguous United States, I show that the proposed method controls the
false positive rate at or near the nominal 5\% level across a wide range
of spatial correlation intensities and sample configurations. I compare
six kernel functions and find that the Bartlett and Epanechnikov kernels
deliver the best size control. An empirical application using U.S.
county-level data illustrates the practical relevance of the method. The
R package \texttt{SpatialInference} implements the proposed bandwidth
selection method.

\vspace{0.5em}


\vspace{0.5em}

\end{abstract}

\setstretch{1.5}
\section{Introduction}\label{sec-intro}

The concept of spatial autocorrelation has served as an organizing
principle in spatial statistics for many decades, dating back at least
to \citet{Student1914}'s article on ``the elimination of spurious
correlation due to position in time and space.'' \citet{Student1914}
demonstrated that the application of conventional statistical methods to
autocorrelated observations ``would lead to altogether misleading
values'' (p.~179) and presented adjusted inferences as evidence of the
importance of such corrections. In the modern econometrics literature,
the problems for statistical inference induced by spatial
autocorrelation have been recognized since at least
\citet{AnselinGriffith1988}, who showed that spatial effects can
substantially distort regression-based inference. Given that virtually
all economic data have spatial structure---and that georeferenced data
are becoming ubiquitous---this is a growing concern in applied research.
The importance of accounting for dependence in inference has been
highlighted in many contexts: \citet{BertrandDufloMullainathan2004}
demonstrated the severity of the problem in panel data with temporal
correlation, and casual empiricism suggests that applied researchers are
now acutely aware of the potential for distortions. Yet for spatial
correlation, no equivalent practical guidance on how to operationalize
the correction has emerged.

The most common approach to address spatial autocorrelation in
regression analysis in economics and many other social sciences is to
rely on spatial heteroskedasticity and autocorrelation consistent (HAC)
standard errors, introduced in the seminal contribution of
\citet{Conley1999}. Conley extended the logic of \citet{NeweyWest1987}
from time-series to the two-dimensional spatial domain, building on the
HAC framework of \citet{Andrews1991}. The core of the technique is a
non-parametric estimator for the variance-covariance matrix that uses
kernel-weighted averages of cross-products of residuals, where the
kernel downweights pairs of observations as a function of the distance
between them. A key input for operationalizing this estimator is the
choice of a \emph{cutoff bandwidth}---the distance beyond which
residuals are assumed to be uncorrelated. In the time-series analogue, a
rich literature on automated and rule-of-thumb bandwidth selection
exists \citep[e.g.,][]{NeweyWest1987, Andrews1991, Sun2014}. For the
spatial case, however, no widely adopted equivalent exists. Instead,
researchers typically rely on ad hoc choices guided by intuition or
domain knowledge, creating what \citet{GelmanLoken2013} term
``researcher degrees of freedom.''

This paper makes three contributions. First, I document an empirical
regularity with important implications for spatial HAC estimation: the
relationship between the bandwidth and the magnitude of spatial HAC
standard errors follows an \emph{inverse-U shape}. This means that both
too narrow and too wide bandwidths lead to underestimated standard
errors and thus to overrejection of true null hypotheses. This finding
contradicts the prevailing conventional wisdom in applied work, which
holds that choosing wider bandwidths ensures conservative inference.

Second, I propose a simple, non-parametric, data-driven method to select
the appropriate bandwidth. The method is based on the empirical
covariogram of regression residuals: the estimated bandwidth corresponds
to the distance at which residual covariation first crosses zero. Third,
I conduct extensive Monte Carlo simulations to evaluate the method.
Using spatial correlation structures calibrated to the geography of the
contiguous United States, I show that the proposed bandwidth selector
controls the false positive rate at or near the nominal 5\% level across
a wide range of spatial autocorrelation intensities, sample sizes, and
spatial configurations. The simulations also provide a systematic
comparison of six commonly used kernel functions---Bartlett, Uniform,
Epanechnikov, Gaussian, Parzen, and Quartic Biweight---for spatial HAC
estimation, finding that the Bartlett and Epanechnikov kernels deliver
the best size control.

\textbf{Related literature.} The most closely related work is
\citet{KimSun2011}, who derive the MSE-optimal bandwidth for spatial HAC
estimation and propose a parametric plug-in estimator. Their optimal
bandwidth is a function of the sample size (\(n^{\alpha}\) for some
\(\alpha > 0\)), which contrasts with the approach proposed here. The
plug-in estimator of \citet{KimSun2011} relies on parametric pilot
estimators of unknown quantities, whereas the covariogram-range method
proposed in this paper is fully non-parametric and does not require such
assumptions. \citet{KelejianPrucha2007} provide a detailed treatment of
the spatial HAC estimator's properties and derive consistency results;
the present paper builds on their theoretical framework.

Recent contributions have taken a different approach by proposing
inference methods that bypass the Conley HAC framework altogether.
\citet{MullerWatson2022} and \citet{MullerWatson2023} develop spatial
correlation principal components (SCPC) confidence intervals that are
robust to arbitrary spatial correlation without requiring bandwidth
selection. \citet{MullerWatson2024} show that spatial I(1) processes can
induce spuriously significant regression results even with spatial HAC
or cluster-robust standard errors, and propose spatial differencing
methods. \citet{ConleyKelly2025} propose spatial basis regressions
combined with placebo tests for persistence studies.
\citet{DellaVignaImbensKimRitzwoller2025} use multiple outcome variables
to estimate cross-sectional dependence, sidestepping the bandwidth
problem entirely. While each of these methods has merit, they also
require departures from the familiar regression framework that applied
researchers are accustomed to. The present paper demonstrates that valid
inference can be achieved \emph{within} the standard Conley HAC
framework---provided the bandwidth is chosen appropriately. The method
is transparent, intuitive, and computationally efficient.

In the time-series HAC literature, \citet{KolokotronesStockWalker2024}
show that the Bartlett (Newey--West) kernel delivers the highest power
among first-order kernels. The present paper extends the comparison of
kernel functions to the spatial domain, where such a systematic
evaluation has not been conducted.

The rest of the paper is organized as follows. Section~\ref{sec-setup}
introduces the econometric framework and the spatial HAC estimator.
Section~\ref{sec-rangecovariogram} presents the proposed bandwidth
selection method with formal assumptions and a consistency result.
Section~\ref{sec-noise} describes the simulation design for generating
spatially correlated data. Section~\ref{sec-inverseu} documents the
inverse-U relationship. Section~\ref{sec-montecarlo} presents the Monte
Carlo results. Section~\ref{sec-empirical} provides an empirical
application. Section~\ref{sec-conclusion} concludes.
Section~\ref{sec-appendix} contains additional details on spatial HAC
estimation, kernel functions, and the geostatistical simulation
procedure.

\section{Econometric Framework}\label{sec-setup}

\subsection{The Linear Model with Spatially Correlated
Errors}\label{the-linear-model-with-spatially-correlated-errors}

Let \(\mathcal{S} \subset \mathbb{R}^2\) be a subset of two-dimensional
Euclidean space and let
\(\mathbf{s}_i = (s_{1i}, s_{2i})' \in \mathcal{S}\) denote the spatial
location of observation \(i\), with \(s_{1i}\) and \(s_{2i}\)
representing the two spatial coordinates.\footnote{For unprojected
  geographic data, these correspond to longitude and latitude on the
  sphere \(\mathbb{S}^2\). For projected data, they are coordinates in a
  planar coordinate system. Without loss of generality, this paper
  refers to locations in \(\mathbb{R}^2\) and implicitly assumes
  projected data.} Consider the linear regression model

\begin{equation}\phantomsection\label{eq-spatial-ols}{
y_i = \mathbf{x}_i' \boldsymbol{\beta} + \varepsilon_i, \quad i = 1, \ldots, n,
}\end{equation}

where \(y_i \equiv y(\mathbf{s}_i)\) is the outcome at location
\(\mathbf{s}_i\), \(\mathbf{x}_i\) is a \(p \times 1\) vector of
regressors, \(\boldsymbol{\beta}\) is the parameter vector of interest,
and \(\varepsilon_i \equiv \varepsilon(\mathbf{s}_i)\) is the error term
with \(E[\varepsilon_i] = 0\). The error terms fail to satisfy the
independence assumption:
\(\text{Cov}(\varepsilon_i, \varepsilon_j) \neq 0\) for nearby locations
\(\mathbf{s}_i\) and \(\mathbf{s}_j\). Intuitively, this arises because
outcomes are not randomly distributed across space; near things tend to
be more related than distant things.\footnote{This paraphrases Tobler's
  \citeyearpar{Tobler1970} first law of geography.}

Since the law of large numbers applies to weakly dependent spatial data,
the OLS estimator \(\hat{\boldsymbol{\beta}}\) remains consistent. The
concern when dealing with spatial autocorrelation is therefore
\emph{inference}: the standard heteroskedasticity-robust variance
estimator \[
\hat{V}_{\text{HC}} = \left( \sum_{i=1}^n \mathbf{x}_i \mathbf{x}_i' \right)^{-1} \left( \sum_{i=1}^n \hat{\varepsilon}_i^2 \, \mathbf{x}_i \mathbf{x}_i' \right) \left( \sum_{i=1}^n \mathbf{x}_i \mathbf{x}_i' \right)^{-1}
\] ignores cross-products \(\hat{\varepsilon}_i \hat{\varepsilon}_j\)
for \(i \neq j\) and thus underestimates the true variance of
\(\hat{\boldsymbol{\beta}}\) in the presence of positive spatial
autocorrelation. This leads to inflated \(t\)-statistics and rejection
rates that exceed the nominal level.

\subsection{The Spatial HAC Estimator}\label{the-spatial-hac-estimator}

\citet{Conley1999} proposes a non-parametric estimator for the
variance-covariance matrix that accounts for spatial correlation. The
spatial HAC estimator takes the form
\begin{equation}\phantomsection\label{eq-vcov-conley}{
\hat{V}_{\text{SHAC}} = \left( \mathbf{X}'\mathbf{X} \right)^{-1} \hat{\mathbf{\Omega}} \left( \mathbf{X}'\mathbf{X} \right)^{-1},
}\end{equation} where \begin{equation}\phantomsection\label{eq-omega}{
\hat{\mathbf{\Omega}} = \sum_{i=1}^n \sum_{j=1}^n K\!\left(\frac{d_{ij}}{\varsigma}\right) \hat{\varepsilon}_i \hat{\varepsilon}_j \, \mathbf{x}_i \mathbf{x}_j'.
}\end{equation}

Here, \(d_{ij} = d(\mathbf{s}_i, \mathbf{s}_j)\) is the distance between
observations \(i\) and \(j\), \(\varsigma > 0\) is the bandwidth (cutoff
distance), and \(K(\cdot)\) is a kernel function satisfying
\(K(0) = 1\), \(K(u) = 0\) for \(|u| > 1\), and \(K(u) = K(-u)\). The
kernel downweights the contribution of cross-products as a function of
distance, assigning zero weight to pairs separated by more than
\(\varsigma\). When \(\varsigma = 0\), the estimator reduces to the
standard heteroskedasticity-consistent (HC) estimator.

The choice of bandwidth \(\varsigma\) is the central concern of this
paper. It determines the distance within which residuals are allowed to
be correlated and thus directly affects the magnitude of the estimated
standard errors.

\subsection{Kernel Functions}\label{sec-kernels-main}

Six kernel functions are considered in this study:

\[
\begin{array}{ll}
\text{Uniform:} & K(u) = \mathbf{1}(0 \leq u \leq 1) \\[4pt]
\text{Bartlett:} & K(u) = (1 - u) \cdot \mathbf{1}(0 \leq u \leq 1) \\[4pt]
\text{Epanechnikov:} & K(u) = (1 - u^2) \cdot \mathbf{1}(0 \leq u \leq 1) \\[4pt]
\text{Parzen:} & K(u) = \begin{cases} 1 - 6u^2 + 6u^3 & 0 \leq u < 1/2 \\ 2(1 - u)^3 & 1/2 \leq u \leq 1 \end{cases} \\[4pt]
\text{Quartic Biweight:} & K(u) = (1 - u^2)^2 \cdot \mathbf{1}(0 \leq u \leq 1) \\[4pt]
\text{Gaussian:} & K(u) = \exp(-u^2/2) \cdot \mathbf{1}(0 < u < 1) \\[4pt]
\end{array}
\]

where \(u = d_{ij}/\varsigma\) is the normalized distance. The Uniform
kernel assigns equal weight to all pairs within the bandwidth, while the
others assign declining weights. The Bartlett kernel corresponds to the
spatial analogue of the \citet{NeweyWest1987} estimator. The Gaussian
kernel used here is truncated at the bandwidth, differing from the
standard Gaussian density by omitting the normalization constant; this
ensures \(K(0) = 1\) and \(K(u) = 0\) for \(u > 1\), consistent with the
spatial HAC framework.

\subsection{Measuring Spatial
Autocorrelation}\label{measuring-spatial-autocorrelation}

In practice, the spatial relations described above are typically
accounted for by a connectivity matrix, often referred to as a weighting
matrix---a concept dating back to at least \citet{Moran1948}. Let
\(\mathbf{W}\) be the \(n \times n\) matrix describing a relationship
between all locations \(\mathbf{s}_1, \ldots, \mathbf{s}_n\), where the
\((i,j)\)-th element, \(w_{ij} \equiv [\mathbf{W}]_{ij}\), refers to the
distance between units \(i\) and \(j\). In general, this distance does
not have to be spatial but can also refer to, e.g., economic or social
distance. If \(w_{ij} \neq 0\), units \(i\) and \(j\) are neighbors. In
practice, \(\mathbf{W}\) will be a sparse matrix in most empirical
applications, meaning that most pairs \(i\) and \(j\) are not classified
as neighbors, i.e., \(w_{ij} = 0\).

To illustrate how a spatially lagged variable is constructed, consider
the \(i\)-th row of the connectivity matrix and \(x_i\), the value of
the scalar variable \(\mathbf{x}\) for unit \(i\). The spatial lag of
\(x_i\) can then be written as \[
[\mathbf{W}\mathbf{x}]_i = w_{i1}x_1 + w_{i2}x_2 + \cdots + w_{in}x_n = \sum_{j=1}^n w_{ij} x_j,
\] which is a weighted sum of the values of the same variable across all
neighbors. In practice, the entries of \(\mathbf{W}\) are usually
row-normalized such that \(\sum_{j=1}^n w_{ij} = 1\) for all rows \(i\).
Note also that the main diagonal is set to zero, \(w_{ii} = 0\), meaning
that a unit is not its own neighbor. Put differently, \(\mathbf{W}\) can
be viewed as the spatial lag operator when operationalized via the
matrix product \(\mathbf{W}\mathbf{x}\).

To test each regression specification for residual autocorrelation, I
rely on Moran's I \citep{Moran1950}. In its basic form it can be written
as \[
I = \frac{n}{\sum_i \sum_j w_{ij}} \cdot \frac{\sum_{i=1}^n \sum_{j=1}^n w_{ij}(\hat{\varepsilon}_i - \bar{\hat{\varepsilon}})(\hat{\varepsilon}_j - \bar{\hat{\varepsilon}})}{\sum_{i=1}^n (\hat{\varepsilon}_i - \bar{\hat{\varepsilon}})^2},
\] where \(n\) is the number of units and \(\sum_i \sum_j w_{ij}\) is
the sum of all spatial weights. The denominator
\(\sum_{i=1}^n (\hat{\varepsilon}_i - \bar{\hat{\varepsilon}})^2\) is
the variance and the numerator involves the spatial lags discussed
above. Note the striking resemblance with Pearson's correlation
coefficient: Moran's I quantifies the degree of correlation of
\(\hat{\varepsilon}_i\) with its spatially lagged neighbors. The
expected value under spatial randomness is \(E[I] = -1/(n-1)\) and
inference is based on the standardized statistic
\(Z = (I - E[I]) / \sqrt{\text{Var}(I)}\). When applied to regression
residuals, the computation accounts for the linear projection following
\citet{Anselin1988} and \citet{CliffOrd1981}. All Moran's I tests are
conducted using the \texttt{spdep} package
\citep{BivandAltmanAnselinAssuncaoBerkeEtAl2020}.

A drawback when working with connectivity matrices is that they usually
have to be defined ex-ante by the researcher. For the present study,
this is less of a concern because the matrix is used only for testing,
not for estimation. The weights matrix is held constant across all
simulations---based on a distance band of 200 km---so that reported
Moran's I values are comparable across scenarios. In practice, there are
many ways to define a connectivity matrix; this study relies on distance
band and \(k\)-nearest neighbor specifications.

\section{The Covariogram Range Method}\label{sec-rangecovariogram}

\subsection{The Empirical Covariogram}\label{the-empirical-covariogram}

The key to the proposed bandwidth selection method is the
\emph{empirical covariogram} of regression residuals. For a set of
estimated residuals \(\hat{\varepsilon}_i\), \(i = 1, \ldots, n\), the
empirical covariogram at lag distance \(h\) is
\begin{equation}\phantomsection\label{eq-empirical-covariogram}{
\hat{C}(h) = \frac{1}{|N(h)|} \sum_{(i,j) \in N(h)} \hat{\varepsilon}_i \, \hat{\varepsilon}_j,
}\end{equation} where
\(N(h) = \{(i,j) : h - \delta \leq d(\mathbf{s}_i, \mathbf{s}_j) < h + \delta\}\)
is the set of all location pairs separated by a distance falling in the
bin centered at \(h\) with half-width \(\delta\), and \(|N(h)|\) is the
cardinality of this set. The bin width \(2\delta\) must be chosen large
enough to ensure sufficient observations per bin. In practice, this
amounts to specifying a number of equally spaced, non-overlapping
distance classes.

The covariogram \(\hat{C}(h)\) estimates the covariance between
residuals as a function of the distance separating them. In the presence
of positive spatial autocorrelation, \(\hat{C}(h)\) is positive for
small \(h\) and declines toward zero as \(h\) increases. Beyond a
certain distance, residuals are approximately uncorrelated and
\(\hat{C}(h)\) fluctuates around zero.

\subsection{The Proposed Bandwidth
Selector}\label{the-proposed-bandwidth-selector}

I propose to use the empirical covariogram to estimate the \emph{range}
of spatial dependence in the regression residuals. This estimated range
then serves as the bandwidth for computing spatial HAC standard errors.

\begin{proposition}[Covariogram Range Estimator]\label{prop:range}
Let $\{h_b\}_{b=1}^B$ denote the bin centers at which the empirical covariogram $\hat{C}(\cdot)$ is evaluated. Define the covariogram range estimator as
\begin{equation}\label{eq-covariogram-range}
\hat{\varsigma} = \min\{h_b : |\hat{C}(h_b)| \leq \eta\},
\end{equation}
where $\eta \geq 0$ is a small tolerance to account for sampling noise. The estimator $\hat{\varsigma}$ identifies the shortest distance at which the magnitude of the estimated residual covariance falls below the tolerance band.
\end{proposition}

In practice, \(\hat{C}(h)\) is evaluated at a fixed set of equally
spaced, non-overlapping distance bins. Setting \(\eta = 0\), the
estimator selects the first bin center at which the empirical
covariogram crosses zero. For covariance functions that decay gradually
rather than reaching zero exactly (see Remark 4 below), a small positive
\(\eta\) can be used; in the Monte Carlo experiments, the default
\(\eta = 0\) performs well because the binned covariogram is
sufficiently noisy that it crosses zero near the effective range.

\begin{remark}
The covariogram and the semivariogram are related by $C(h) = C(0) - \gamma(h)$, where $C(0)$ is the variance at distance zero (the "nugget" plus "sill" in geostatistical terminology) and $\gamma(h)$ is the semivariogram. Both convey equivalent information about the spatial correlation structure. The covariogram is used here because the range---the distance at which $C(h)$ crosses zero---has a direct interpretation as the bandwidth beyond which residuals are uncorrelated.
\end{remark}

\subsection{Formal Assumptions and
Consistency}\label{formal-assumptions-and-consistency}

This section provides sufficient conditions under which the covariogram
range estimator \(\hat{\varsigma}\) is consistent for the true
correlation range and the resulting plug-in spatial HAC estimator is
consistent for the variance of \(\hat{\boldsymbol{\beta}}\). The
conditions are stated at a level that makes transparent what is
required; a fully detailed proof in the increasing-domain asymptotic
framework is beyond the scope of this paper. The Monte Carlo evidence in
Section~\ref{sec-montecarlo} provides the primary validation that the
method works in finite-sample settings calibrated to empirically
relevant spatial configurations.

\begin{assumption}[Stationarity and Mixing]\label{ass:stationarity}
The error process $\{\varepsilon(\mathbf{s}) : \mathbf{s} \in \mathcal{S}\}$ is second-order stationary: $E[\varepsilon(\mathbf{s})] = 0$ for all $\mathbf{s}$, and $\text{Cov}(\varepsilon(\mathbf{s}_i), \varepsilon(\mathbf{s}_j)) = C(\|\mathbf{s}_i - \mathbf{s}_j\|)$ depends only on the distance $\|\mathbf{s}_i - \mathbf{s}_j\|$ (isotropy). The process satisfies a spatial $\alpha$-mixing condition with mixing coefficients $\alpha(k)$ satisfying $\sum_{k=1}^{\infty} k^{d-1} \alpha(k)^{\delta/(4+\delta)} < \infty$ for dimension $d = 2$. Additionally, $E[|\varepsilon(\mathbf{s})|^{4+\delta}] < \infty$ for some $\delta > 0$.
\end{assumption}

The mixing and moment conditions are needed because pairs of residual
products \(\hat{\varepsilon}_i \hat{\varepsilon}_j\) sharing common
locations are dependent; they ensure a law of large numbers for the
binned covariogram averages, which does not follow from the growth in
the number of pairs \(|N(h)|\) alone.

\begin{assumption}[Compact Support and Identification]\label{ass:finiterange}
There exists a finite $\varsigma_0 > 0$ such that $C(h) = 0$ for all $h > \varsigma_0$. The covariance function satisfies the separation condition:
\begin{equation}\label{eq-separation}
\text{for every } \delta > 0, \quad \inf_{h \leq \varsigma_0 - \delta} |C(h)| > 0 \quad \text{and} \quad \sup_{h \geq \varsigma_0 + \delta} |C(h)| = 0.
\end{equation}
Additionally, $C(h)$ is continuous at $\varsigma_0$.
\end{assumption}

The separation condition ensures that the true range \(\varsigma_0\) is
well-identified: the covariance is bounded away from zero inside the
range and exactly zero outside. This rules out pathological cases where
\(C(h)\) is ``flat'\,' near \(\varsigma_0\), which would make the
threshold crossing unstable under small estimation errors. Note that
\(C(h)\) need not be positive for all \(h < \varsigma_0\)---it may be
negative at some distances (e.g., due to oscillatory behavior or
residualization)---but it must be nonzero.

\begin{assumption}[Increasing-Domain Sampling]\label{ass:sampling}
The observation locations $\{\mathbf{s}_1, \ldots, \mathbf{s}_n\}$ are sampled in an increasing-domain framework: the spatial region $\mathcal{S}_n$ expands as $n \to \infty$, with minimum inter-point separation bounded away from zero. The number of pairs $|N(h)|$ in each distance bin grows to infinity for all $h \leq \varsigma_0 + c$ for some $c > 0$.
\end{assumption}

This is consistent with the asymptotic framework of \citet{Conley1999}
and \citet{KelejianPrucha2007}, where spatial HAC consistency is
established under increasing-domain asymptotics. The condition that
\(|N(h)| \to \infty\) across the relevant range of distances ensures
that the binned covariogram averages converge.

\begin{assumption}[Regularity]\label{ass:regularity}
The regressors $\mathbf{x}_i$ are bounded, and the conditions of @Conley1999 and @KelejianPrucha2007 for the consistency of the spatial HAC estimator are satisfied with bandwidth $\varsigma_0$. The kernel $K(\cdot)$ is continuous and has compact support (i.e., it is one of the smooth kernels defined in @sec-kernels-main, such as Bartlett or Epanechnikov).
\end{assumption}

The kernel continuity requirement ensures that the mapping from
bandwidth to the SHAC variance estimate,
\(\varsigma \mapsto \hat{V}_{\text{SHAC}}(\varsigma)\), is continuous:
adding or removing pairs near the cutoff boundary produces only a
negligible change in the estimate, because their kernel weights are near
zero.

\begin{proposition}[Consistency of the Covariogram Range Estimator]\label{prop:consistency}
Under Assumptions \ref{ass:stationarity}--\ref{ass:regularity}, the covariogram range estimator $\hat{\varsigma}$ converges in probability to $\varsigma_0$ as $n \to \infty$. Consequently, the spatial HAC estimator $\hat{V}_{\text{SHAC}}$ computed with bandwidth $\hat{\varsigma}$ is consistent for the true variance of $\hat{\boldsymbol{\beta}}$.
\end{proposition}

\emph{Sketch of argument.} Let \(\{h_b\}_{b=1}^B\) be the fixed bin
centers covering \([0, \varsigma_0 + c]\). Under Assumptions
\ref{ass:stationarity} and \ref{ass:sampling}, the mixing and moment
conditions ensure a law of large numbers for the dependent pairwise
averages within each bin, yielding \(\hat{C}(h_b) \to_p C(h_b)\) for
each \(b\). Since \(B\) is finite, joint convergence holds, and the
maximum over bins converges: \[
\max_{b \leq B} |\hat{C}(h_b) - C(h_b)| \to_p 0.
\] Because \(\hat{C}(h)\) is defined as a piecewise-constant function on
the bins, this is equivalent to uniform convergence over the interval.
Given this and the separation condition in Assumption
\ref{ass:finiterange}, the estimator \(\hat{\varsigma}\) converges to
\(\varsigma_0\): for any \(\delta > 0\), eventually \(|\hat{C}(h_b)|\)
is bounded away from zero (and hence above \(\eta\)) for all bins
\(h_b \leq \varsigma_0 - \delta\), and \(|\hat{C}(h_b)|\) is close to
zero (and hence below \(\eta\)) for all bins
\(h_b \geq \varsigma_0 + \delta\). Thus the first bin at which
\(|\hat{C}(h_b)| \leq \eta\) must fall in
\((\varsigma_0 - \delta, \varsigma_0 + \delta)\), giving
\(\hat{\varsigma} \to_p \varsigma_0\). The second claim follows from the
kernel continuity in Assumption \ref{ass:regularity}---which ensures
\(\hat{V}_{\text{SHAC}}(\hat{\varsigma}) - \hat{V}_{\text{SHAC}}(\varsigma_0) \to_p 0\).
Specifically, under increasing-domain sampling, the number of
observation pairs in a thin boundary shell
\((\varsigma_0 - \epsilon, \varsigma_0 + \epsilon)\) grows at most
proportionally to \(\epsilon\), so with a continuous compact-support
kernel the marginal contribution of these pairs to
\(\hat{\mathbf{\Omega}}\) vanishes as \(\epsilon \downarrow 0\).
Combined with the consistency of \(\hat{V}_{\text{SHAC}}(\varsigma_0)\)
established by \citet{KelejianPrucha2007}, this yields the result.

\begin{remark}
The stationarity assumption rules out spatial trends and non-stationary spatial processes. In practice, non-stationarity can be addressed by including location controls (e.g., coordinates as regressors) or by detrending before applying the covariogram method. The Monte Carlo experiments in @sec-montecarlo include specifications both with and without location controls.
\end{remark}

\begin{remark}
The compact support assumption (Assumption \ref{ass:finiterange}) is a modeling approximation. It is exactly satisfied by the Spherical covariance model but not by the Exponential or Mat\'{e}rn, whose covariances decay to zero asymptotically without reaching it in finite distance. In practice, these models have an ``effective range''---the distance at which the covariance falls below approximately 5\% of its sill value---and the covariogram estimator targets this effective range. Formally, one could replace $\varsigma_0$ with the pseudo-true parameter $\varsigma_0 := \sup\{h : |C(h)| > \eta\}$ for a small threshold $\eta > 0$, and the argument above would carry through with the separation condition stated relative to this threshold. Under compact support, the SHAC estimator is consistent at the true cutoff $\varsigma_0$, and the covariogram range provides a consistent estimate of this cutoff. Under decaying covariance without compact support, the covariogram range targets a fixed effective range that serves as a practical bandwidth; formal rate-optimality in that regime is left for future work.
\end{remark}

\subsection{Implementation}\label{implementation}

The proposed method is computationally efficient. The empirical
covariogram is computed using the \texttt{variogram()} function from
\texttt{gstat} \citep{Pebesma2004} with the
\texttt{covariogram\ =\ TRUE} option. The distance bins are determined
by two parameters: \texttt{width} (the bin width \(2\delta\)) and
\texttt{cutoff} (the maximum distance). As a default, the bin width is
set to ensure approximately 100--200 distance bins up to two-thirds of
the maximum inter-point distance. The range is then extracted as the
first distance at which the empirical covariogram crosses zero. For a
typical dataset with several thousand observations, the entire procedure
takes a fraction of a second.

Figure~\ref{fig-covgillustr} illustrates the method on four simulated
datasets with increasing degrees of spatial autocorrelation.

\subsection{\texorpdfstring{Connection to Fixed-\(b\)
Asymptotics}{Connection to Fixed-b Asymptotics}}\label{sec-fixedb}

The covariogram-range bandwidth selector has a useful conceptual
connection to the fixed-\(b\) asymptotic framework. In the classical
time-series setting, two asymptotic regimes govern the behavior of HAC
standard errors. Under the traditional \emph{small-\(b\)} asymptotics of
\citet{Andrews1991} and \citet{NeweyWest1987}, the bandwidth \(M\) grows
with the sample size \(T\) but satisfies \(M/T \to 0\), so that the HAC
estimator is consistent and the \(t\)-statistic converges to a standard
normal. Under the \emph{fixed-\(b\)} asymptotics of
\citet{KieferVogelsangBunzel2000} and \citet{SunPhillipsJin2008}, the
bandwidth satisfies \(M/T \to b \in (0, 1]\), so the HAC estimator is
inconsistent but the resulting test statistic converges to a
non-standard distribution that provides more accurate finite-sample size
control. \citet{Sun2014} argues that fixed-\(b\) inference is generally
preferable, as it accounts for the estimation uncertainty in the HAC
estimator itself. \citet{BesterConleyHansenVogelsang2016} extend
fixed-\(b\) theory to the spatial setting: when the Conley bandwidth
grows proportionally with the diameter of the spatial domain---i.e.,
\(\varsigma / \text{diam}(S_n) \to b > 0\)---the SHAC estimator
converges to a nondegenerate random matrix rather than the true
variance, and the \(t\)-statistic has a non-standard pivotal limit
distribution whose critical values must be obtained by simulation.

While the covariogram-range selector does not implement spatial
fixed-\(b\) inference, it shares the key practical intuition that
motivates the fixed-\(b\) literature: smoothing should be anchored to
the scale of dependence rather than shrinking relative to the sample.
The estimator \(\hat{\varsigma}\) targets a fixed physical
quantity---the correlation range \(\varsigma_0\)---determined entirely
by the spatial dependence structure of the data-generating process.
Unlike MSE-optimal rules \citep[e.g.,][]{Andrews1991}, where the
bandwidth grows with \(T\) but satisfies \(M/T \to 0\) and tends to
undersmooth in finite samples, the covariogram-range bandwidth does not
depend on \(n\) and is instead pinned to the dependence scale. This
fixed-span behavior---anchoring to a physical quantity rather than to
sample size---reduces the usual smoothing trade-off that \citet{Sun2014}
identifies as the primary source of size distortions in HAC inference.

Formally, however, the covariogram-range method operates in a different
asymptotic regime. Under the increasing-domain asymptotics used
throughout this paper, a fixed physical bandwidth implies
\(\varsigma_0 / \text{diam}(S_n) \to 0\), placing the covariogram-range
selector in the small-\(b\) regime of
\citet{BesterConleyHansenVogelsang2016}, not the fixed-\(b\) regime.
Moreover, under compact support (Assumption \ref{ass:finiterange}), the
covariance is literally zero beyond \(\varsigma_0\), so there is no
``missing tail'' to truncate---unlike the spatial fixed-\(b\) setting of
\citet{BesterConleyHansenVogelsang2016}, where the bandwidth truncates a
non-negligible portion of the dependence structure and the SHAC
estimator converges to a nondegenerate random limit. At the true cutoff
\(\varsigma_0\), the SHAC estimator captures the entire covariance
structure without truncation bias and is therefore consistent. The
plug-in bandwidth \(\hat{\varsigma}\) converges to \(\varsigma_0\),
standard SHAC consistency obtains, and no non-standard critical values
or simulated reference distributions are required. The finite-sample
size improvements documented in Section~\ref{sec-montecarlo} arise not
from the fixed-\(b\) mechanism of embracing estimator randomness, but
from targeting the correct dependence scale directly---avoiding the
undersmoothing that plagues MSE-optimal bandwidth selectors.
Nevertheless, in any finite spatial sample the bandwidth-to-domain ratio
is a non-negligible constant, so the resulting smoothing can resemble
the larger-smoothing behavior that motivates fixed-\(b\) corrections.

\section{Simulation Design}\label{sec-noise}

To evaluate the proposed bandwidth selector, I require spatially
correlated random variables with precisely controlled correlation
ranges. I generate these using unconditional geostatistical simulation
(simple kriging) with the \texttt{gstat} package \citep{Pebesma2004};
see \citet{Cressie1993} for a comprehensive treatment of the
geostatistical framework. The full details of the simulation
procedure---including the kriging system, semivariogram specifications,
and implementation---are provided in Section~\ref{sec-appendix-kriging}.

\subsection{Spatial Samples}\label{spatial-samples}

Four spatial samples are used throughout the study, all covering the
bounding box of the contiguous United States:

\begin{enumerate}
\def\labelenumi{\arabic{enumi}.}
\tightlist
\item
  \textbf{Counties} (\(n = 3{,}108\)): Centroids of all counties in the
  contiguous 48 states. This is a sample with highly unequal spacing
  between observations.
\item
  \textbf{Coarse lattice} (\(n \approx 2{,}600\)): A regular square grid
  with 70 km cell size.
\item
  \textbf{Medium lattice} (\(n \approx 5{,}096\)): A regular square grid
  with 50 km cell size.
\item
  \textbf{Fine lattice} (\(n \approx 10{,}320\)): A regular square grid
  with 35 km cell size.
\end{enumerate}

The three lattice samples provide controlled variation in sample
density, while the county centroid sample tests performance under highly
irregular spacing.

\subsection{Monte Carlo Design}\label{monte-carlo-design}

In each simulation iteration:

\begin{enumerate}
\def\labelenumi{\arabic{enumi}.}
\item
  Two independent spatially correlated random fields are drawn using the
  procedure described above and in Section~\ref{sec-appendix-kriging},
  both with the same range parameter. The first provides the dependent
  variable \(y_i\) (labeled \texttt{noise1}) and the second provides the
  regressor \(x_i\) (labeled \texttt{noise2}).
\item
  Each observation is assigned the value of the underlying random field
  at its location.
\item
  A univariate regression of \(y_i\) on \(x_i\) is estimated. Because
  the two fields are drawn independently, the true coefficient is zero
  (\(\beta = 0\)). The \(t\)-statistics using HC1 standard errors and
  using Conley standard errors with various kernel and bandwidth
  combinations are recorded.
\item
  The false positive rate is defined as the fraction of simulations in
  which the null hypothesis \(H_0: \beta = 0\) is rejected at the 5\%
  level using the critical value of 1.96.
\end{enumerate}

This process is repeated for noise range parameters starting from 0 (no
spatial correlation) in 15 sequential increments until extreme degrees
of spatial correlation are reached, with 5,000 iterations per
configuration. The noise range parameter governs the input to the
semivariogram; the \emph{realized} correlation range varies
stochastically across draws. Computations were carried out on the
Akropolis cluster at the University of Chicago using \texttt{Rmpi} for
parallel processing \citep{RCoreTeam2025}. To ensure exact
replicability, the pseudo-random number seed was set using
\texttt{clusterSetRNGStream()} with Pierre L'Ecuyer's
\texttt{RngStreams} \citep{LEcuyer1999} (seed: 1908).

For each simulation, the following standard errors are computed:

\begin{itemize}
\tightlist
\item
  \textbf{HC1}: The standard heteroskedasticity-robust estimator
  (benchmark).
\item
  \textbf{Conley, covariogram-range bandwidth} (\(\hat{\varsigma}\)):
  The proposed method, computed with six different kernels.
\item
  \textbf{Conley, fixed narrow bandwidth} (25 km): A deliberately small
  bandwidth.
\item
  \textbf{Conley, fixed wide bandwidth} (2,500 km): A deliberately large
  bandwidth representing the ``conservative wide bandwidth'' advice.
\end{itemize}

\section{The Inverse-U Relationship}\label{sec-inverseu}

\begin{figure}

\begin{minipage}{0.25\linewidth}

\centering{

\includegraphics{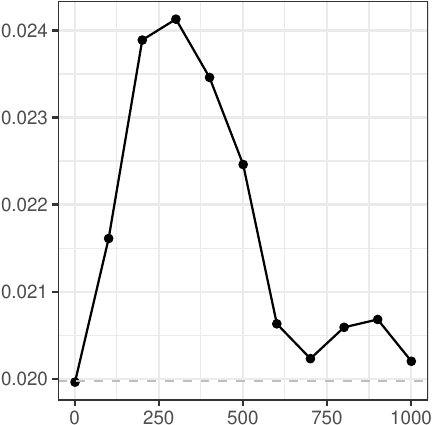}

}

\subcaption{\label{fig-inverseu-1}Coarse lattice
(\(n \approx 2{,}600\)).}

\end{minipage}
\begin{minipage}{0.25\linewidth}

\centering{

\includegraphics{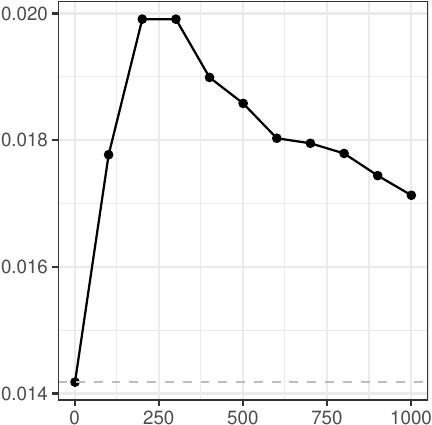}

}

\subcaption{\label{fig-inverseu-2}Medium lattice
(\(n \approx 5{,}096\)).}

\end{minipage}
\begin{minipage}{0.25\linewidth}

\centering{

\includegraphics{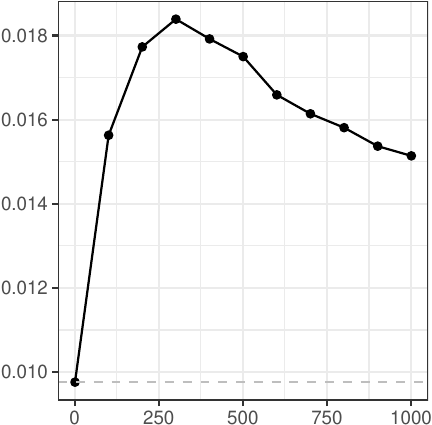}

}

\subcaption{\label{fig-inverseu-3}Fine lattice
(\(n \approx 10{,}320\)).}

\end{minipage}
\begin{minipage}{0.25\linewidth}

\centering{

\includegraphics{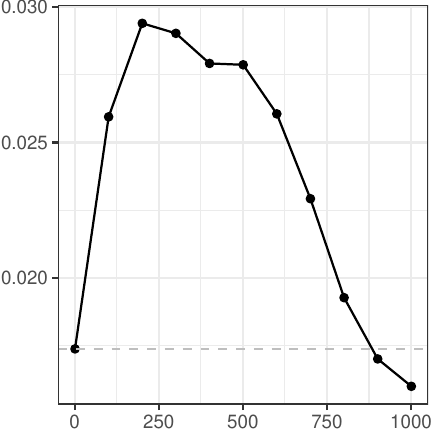}

}

\subcaption{\label{fig-inverseu-4}County centroids (\(n = 3{,}108\)).}

\end{minipage}

\caption{\label{fig-inverseu}The inverse-U relationship between the
magnitude of spatial HAC standard errors and the kernel bandwidth for
all four samples. Each panel uses a single realization of a spatially
correlated random field (Mat\textquotesingle ern, range 50 km). The
dashed grey line indicates the corresponding HC1 standard error. It can
be seen that overly wide bandwidths drive the size of the Conley
standard error down to the levels of the HC1 standard error with overly
inflated t-stats.}

\end{figure}

Figure~\ref{fig-inverseu} displays the relationship between the kernel
bandwidth and the magnitude of Conley standard errors for a single
realization of a spatially correlated random field across the four
spatial samples. The pattern is striking and consistent: the standard
error first \emph{increases} as the bandwidth grows from zero (at which
point the Conley SE equals the HC1 SE), reaches a maximum near the true
correlation range, and then \emph{decreases} steadily. The method
proposed in this paper always picks the peak of this inverse-U curve as
the standard error.

This inverse-U shape has immediate practical consequences. Starting from
a narrow bandwidth, expanding the range over which residuals are allowed
to be correlated initially leads to larger standard errors---as
expected, since positive residual correlation adds to the estimated
variance. However, beyond the range of actual correlation, additional
observations included in the kernel contribute only noise to the
variance estimate: their cross-products
\(\hat{\varepsilon}_i \hat{\varepsilon}_j\) have expectation zero but
dilute the positive contributions from truly correlated pairs. As the
bandwidth continues to grow, this dilution effect dominates and the
standard error declines.

For very wide bandwidths, the standard error can fall \emph{below} the
HC1 level. This directly contradicts the conventional wisdom that wider
bandwidths ensure conservative inference. In fact, choosing a bandwidth
much larger than the true correlation range can lead to standard errors
that are \emph{smaller} than even the heteroskedasticity-only estimator,
resulting in overrejection that exceeds even the overrejection from
ignoring spatial correlation entirely.

The practical implication is clear: the bandwidth should be matched to
the actual range of spatial correlation in the residuals. This is
precisely what the covariogram range method proposed in
Section~\ref{sec-rangecovariogram} achieves.

\section{Illustrating the Covariogram Range
Method}\label{sec-illustration}

\begin{figure}

\begin{minipage}{0.33\linewidth}

\centering{

\includegraphics{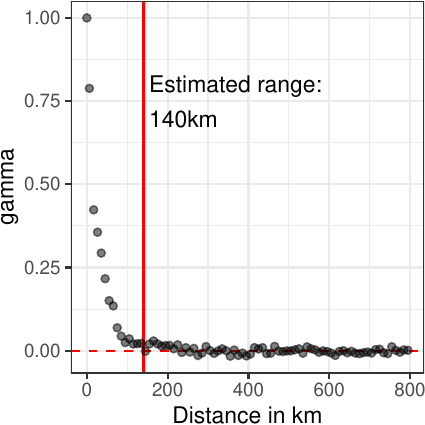}

}

\subcaption{\label{fig-covgillustr-1}Low SAC}

\end{minipage}
\begin{minipage}{0.33\linewidth}

\centering{

\includegraphics{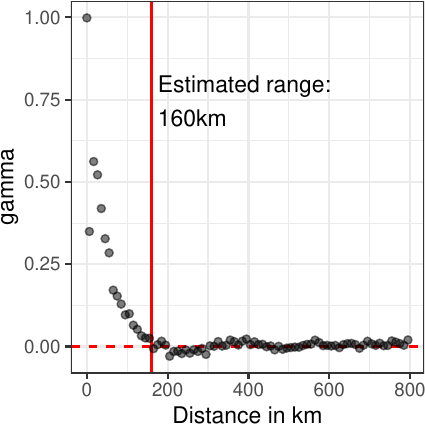}

}

\subcaption{\label{fig-covgillustr-2}Moderate SAC}

\end{minipage}
\begin{minipage}{0.33\linewidth}

\centering{

\includegraphics{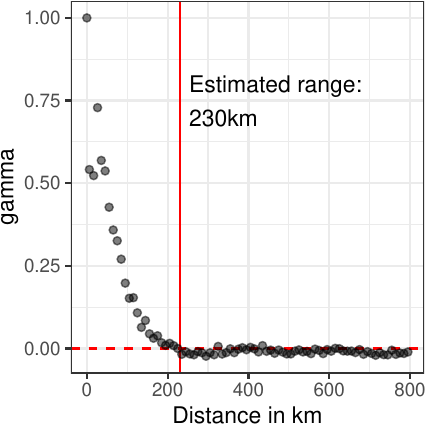}

}

\subcaption{\label{fig-covgillustr-3}High SAC}

\end{minipage}

\caption{\label{fig-covgillustr}Empirical covariograms of regression
residuals from univariate regressions of two independently drawn spatial
noise fields, estimated on 3,108 county centroids. The red vertical line
indicates the estimated correlation range \(\hat{\varsigma}\). This
corresponds to the peak of the inverse-U curve illustrated earlier. The
red dashed horizontal line marks zero covariance.}

\end{figure}

Figure~\ref{fig-covgillustr} visualizes the covariogram range method on
three regression residual series with increasing degrees of spatial
autocorrelation. In each panel, the empirical covariogram declines from
a positive value at short distances and crosses zero at the estimated
range \(\hat{\varsigma}\) (marked by the red vertical line). Beyond
\(\hat{\varsigma}\), the covariogram fluctuates around zero, indicating
the absence of residual correlation. As the degree of spatial
autocorrelation increases from panel (a) to (c), the estimated range
shifts rightward, correctly tracking the growing extent of spatial
dependence.

\section{Monte Carlo Results}\label{sec-montecarlo}

\subsection{Main Results: Size Control}\label{main-results-size-control}

\begin{table}

\centering{

[!h]
\centering\begingroup\fontsize{9}{11}\selectfont

\resizebox{\ifdim\width>\linewidth\linewidth\else\width\fi}{!}{
\begin{tabular}[t]{ccc>{}ccc>{}ccc>{}c}
\toprule
\multicolumn{1}{c}{ } & \multicolumn{3}{c}{$n = 2{,}600$ (70km grid)} & \multicolumn{3}{c}{$n = 5{,}096$ (50km grid)} & \multicolumn{3}{c}{$n = 10{,}320$ (35km grid)} \\
\cmidrule(l{3pt}r{3pt}){2-4} \cmidrule(l{3pt}r{3pt}){5-7} \cmidrule(l{3pt}r{3pt}){8-10}
Noise \# & Range & HC1 & Epan. & Range & HC1 & Epan. & Range & HC1 & Epan.\\
\midrule
0 & 143 & 5.2 & \textbf{5.2} & 109 & 4.9 & \textbf{5.1} & 83 & 4.8 & \textbf{4.9}\\
1 & 141 & 4.6 & \textbf{4.7} & 98 & 4.5 & \textbf{4.5} & 83 & 4.7 & \textbf{4.7}\\
2 & 131 & 4.9 & \textbf{4.9} & 140 & 5.0 & \textbf{4.8} & 150 & 7.1 & \textbf{5.3}\\
3 & 178 & 6.1 & \textbf{5.7} & 209 & 7.2 & \textbf{5.5} & 226 & 13.6 & \textbf{6.1}\\
4 & 231 & 7.1 & \textbf{5.6} & 267 & 12.3 & \textbf{6.1} & 295 & 21.5 & \textbf{5.8}\\
\addlinespace
5 & 284 & 9.2 & \textbf{5.5} & 324 & 17.0 & \textbf{6.0} & 355 & 29.9 & \textbf{6.3}\\
6 & 335 & 13.7 & \textbf{6.4} & 370 & 23.2 & \textbf{6.4} & 406 & 38.9 & \textbf{6.5}\\
7 & 382 & 17.7 & \textbf{6.5} & 421 & 29.6 & \textbf{6.4} & 451 & 46.3 & \textbf{6.6}\\
8 & 429 & 22.0 & \textbf{6.9} & 468 & 36.2 & \textbf{6.9} & 500 & 50.8 & \textbf{6.7}\\
9 & 472 & 25.6 & \textbf{6.6} & 509 & 40.4 & \textbf{6.8} & 537 & 54.8 & \textbf{7.2}\\
\addlinespace
10 & 510 & 30.6 & \textbf{7.4} & 551 & 45.5 & \textbf{7.8} & 581 & 60.0 & \textbf{7.5}\\
11 & 540 & 35.5 & \textbf{7.9} & 579 & 49.1 & \textbf{8.4} & 604 & 62.8 & \textbf{8.3}\\
12 & 576 & 38.8 & \textbf{8.0} & 619 & 51.4 & \textbf{8.5} & 647 & 65.5 & \textbf{8.4}\\
13 & 614 & 42.0 & \textbf{8.7} & 654 & 55.7 & \textbf{8.7} & 678 & 68.0 & \textbf{8.5}\\
14 & 639 & 45.2 & \textbf{9.1} & 679 & 57.9 & \textbf{9.0} & 702 & 69.4 & \textbf{8.9}\\
\addlinespace
15 & 660 & 47.7 & \textbf{8.5} & 701 & 60.0 & \textbf{8.8} & 722 & 71.6 & \textbf{9.2}\\
\bottomrule
\end{tabular}}
\endgroup{}

}

\caption{\label{tbl-main}Null rejection frequencies (\%) of nominal 5\%
level tests for linear regressions using the three regular lattice
samples. The table juxtaposes HC1 standard errors with spatial HAC
standard errors using the introduced covariogram-range bandwidth and an
Epanechnikov kernel. Results based on 5,000 Monte Carlo simulations for
every range-sample pair. The range column illustrates the increasing
degree of spatial correlation as the range of the simulated spatial
fields increases. As can be seen from the overly inflated HC1 rejection
rates, the last few rows of the table showcase unrealistically high
scenarios of spatial correlation. Despite that, the proposed method
reduces a rejection rate of over 70\% to around 9\%.}

\end{table}

Table~\ref{tbl-main} presents the main results across the three regular
lattice samples. The covariogram-range method with the Epanechnikov
kernel consistently maintains rejection rates near 5\% across all sample
sizes and spatial configurations, while HC1 rejection rates rise sharply
with spatial autocorrelation. Larger samples (\(n = 10{,}320\)) tend to
produce slightly more accurate size control, consistent with the
consistency result in Proposition \ref{prop:consistency}.

\begin{table}

\centering{

[!h]
\centering\begingroup\fontsize{9}{11}\selectfont

\begin{tabular}[t]{ccccc>{}c>{}c>{}c}
\toprule
\multicolumn{3}{c}{ } & \multicolumn{2}{c}{Fixed bandwidth} & \multicolumn{3}{c}{Covariogram range} \\
\cmidrule(l{3pt}r{3pt}){4-5} \cmidrule(l{3pt}r{3pt}){6-8}
Noise \# & Range (km) & HC1 & Fixed 25km & Fixed 2500km & Bartlett & Uniform & Epan.\\
\midrule
0 & 49 & 4.8 & 4.7 & 24.4 & \textbf{4.8} & \textbf{5.1} & \textbf{4.9}\\
1 & 72 & 5.1 & 5.0 & 23.6 & \textbf{4.9} & \textbf{5.0} & \textbf{4.9}\\
2 & 130 & 6.7 & 6.3 & 24.6 & \textbf{5.6} & \textbf{5.5} & \textbf{5.4}\\
3 & 189 & 11.6 & 10.5 & 25.5 & \textbf{6.4} & \textbf{6.5} & \textbf{6.1}\\
4 & 243 & 18.1 & 16.3 & 26.8 & \textbf{7.8} & \textbf{7.5} & \textbf{7.2}\\
\addlinespace
5 & 287 & 25.2 & 23.0 & 27.8 & \textbf{8.3} & \textbf{7.9} & \textbf{6.9}\\
6 & 324 & 32.8 & 30.4 & 29.4 & \textbf{9.3} & \textbf{9.0} & \textbf{7.7}\\
7 & 360 & 37.1 & 34.7 & 30.2 & \textbf{9.6} & \textbf{9.3} & \textbf{8.2}\\
8 & 386 & 43.5 & 40.5 & 30.1 & \textbf{10.8} & \textbf{10.1} & \textbf{9.1}\\
9 & 418 & 47.3 & 44.4 & 30.1 & \textbf{11.3} & \textbf{11.2} & \textbf{9.5}\\
\addlinespace
10 & 445 & 52.7 & 49.8 & 30.4 & \textbf{11.5} & \textbf{11.0} & \textbf{9.7}\\
11 & 464 & 56.6 & 54.2 & 31.9 & \textbf{13.0} & \textbf{12.3} & \textbf{11.0}\\
12 & 484 & 58.0 & 55.5 & 30.7 & \textbf{13.2} & \textbf{13.0} & \textbf{11.5}\\
13 & 506 & 61.2 & 59.0 & 31.6 & \textbf{14.2} & \textbf{13.2} & \textbf{12.0}\\
14 & 523 & 64.5 & 62.2 & 31.9 & \textbf{14.9} & \textbf{14.6} & \textbf{12.8}\\
\addlinespace
15 & 529 & 64.4 & 62.0 & 32.1 & \textbf{14.6} & \textbf{14.2} & \textbf{12.1}\\
\bottomrule
\end{tabular}
\endgroup{}

}

\caption{\label{tbl-county}Null rejection frequencies (\%) of nominal
5\% level tests for linear regressions for the county centroid sample
(\(n = 3{,}108\)). The table juxtaposes HC1 standard errors with spatial
HAC standard errors using the introduced covariogram-range bandwidth
(showcasing the rejection frequencies of the Bartlett, Uniform, and
Epanechnikov kernel next to each other). The fixed bandwidth columns
illustrate the failure of both too narrow and too wide cutoff bandwidths
for the spatial HAC standard error estimate after Conley (1999). Results
based on 5,000 Monte Carlo simulations for every noise range grouping.
The range column illustrates the increasing degree of spatial
correlation as the range of the simulated spatial fields increases. As
can be seen from the overly inflated HC1 rejection rates, the last few
rows of the table showcase unrealistically high scenarios of spatial
correlation.}

\end{table}

Table~\ref{tbl-county} presents the detailed results for the county
centroid sample (\(n = 3{,}108\)). Several findings stand out.

First, HC1 standard errors lead to substantial overrejection as spatial
autocorrelation increases. At the highest noise ranges, the false
positive rate exceeds 60\%, far above the nominal 5\%.

Second, a fixed narrow bandwidth of 25 km offers no improvement over HC1
when the true correlation range exceeds this bandwidth. Conversely, a
fixed wide bandwidth of 2,500 km---meant to be ``conservative''---also
fails to control size, confirming the inverse-U prediction: at very wide
bandwidths, the standard error is deflated below the HC1 level.

Third, the proposed covariogram-range method with Bartlett, Uniform, or
Epanechnikov kernels controls the rejection rate remarkably well. Across
all noise range configurations, the false positive rate remains close to
the nominal 5\%. At the most extreme degrees of spatial autocorrelation
(with noise ranges far above empirically realistic levels), the
rejection rate rises to approximately 8--10\%, reflecting the inherent
difficulty of spatial HAC estimation when the correlation range
approaches a substantial fraction of the spatial domain.

Taken together, the results demonstrate that the covariogram-range
bandwidth selector provides effective size control across the
empirically relevant spectrum of spatial dependence. For the correlation
ranges most commonly encountered in applied work---where HC1 rejection
rates already exceed 15--30\%---the proposed method maintains false
positive rates in the range of 5--8\%, close to the nominal level. The
modest overrejection observed at extreme correlation ranges reflects a
regime that pushes far beyond what typical georeferenced datasets
exhibit, and even there the covariogram-range method substantially
outperforms all fixed-bandwidth alternatives.

\subsection{Kernel Comparison}\label{kernel-comparison}

\begin{table}

\centering{

[!h]
\centering\begingroup\fontsize{9}{11}\selectfont

\begin{tabular}[t]{ccccccc}
\toprule
Noise \# & Bartlett & Uniform & Epan. & Gaussian & Parzen & Biweight\\
\midrule
0 & 4.9 & 5.0 & 5.1 & 5.0 & 4.8 & 5.0\\
1 & 4.5 & 4.8 & 4.5 & 4.7 & 4.5 & 4.5\\
2 & 4.9 & 5.1 & 4.8 & 5.0 & 4.9 & 4.9\\
3 & 5.9 & 5.6 & 5.5 & 5.6 & 5.9 & 5.6\\
4 & 6.8 & 6.2 & 6.1 & 6.2 & 7.0 & 6.3\\
\addlinespace
5 & 7.2 & 6.0 & 6.0 & 6.0 & 7.5 & 6.4\\
6 & 7.6 & 6.7 & 6.4 & 6.6 & 7.8 & 6.6\\
7 & 8.0 & 7.0 & 6.4 & 6.6 & 8.4 & 6.7\\
8 & 8.7 & 7.3 & 6.9 & 7.1 & 8.7 & 7.5\\
9 & 8.5 & 7.3 & 6.8 & 7.1 & 8.5 & 7.2\\
\addlinespace
10 & 9.4 & 8.2 & 7.8 & 7.9 & 9.6 & 7.9\\
11 & 10.2 & 9.0 & 8.4 & 8.6 & 10.5 & 8.6\\
12 & 10.4 & 9.2 & 8.5 & 8.7 & 10.7 & 8.7\\
13 & 10.8 & 9.8 & 8.7 & 9.4 & 10.9 & 8.9\\
14 & 10.8 & 10.3 & 9.0 & 9.9 & 10.8 & 8.9\\
\addlinespace
15 & 10.9 & 10.0 & 8.8 & 9.7 & 11.3 & 9.1\\
\bottomrule
\end{tabular}
\endgroup{}

}

\caption{\label{tbl-kernels}Null rejection frequencies (\%) by kernel
function using the covariogram-range bandwidth. Regular lattice sample
(\(n = 5{,}096\), 50km grid). Results based on 5,000 simulations per
configuration.}

\end{table}

Table~\ref{tbl-kernels} reports rejection frequencies for all six kernel
functions. The Bartlett and Epanechnikov kernels consistently deliver
rejection rates closest to the nominal 5\% level. The Uniform kernel
performs well at moderate spatial autocorrelation but tends to slightly
overreject at high levels. The Gaussian, Parzen, and Quartic Biweight
kernels show somewhat more variable performance, with a tendency toward
overrejection at higher noise ranges. Based on these results, I
recommend the \textbf{Bartlett} or \textbf{Epanechnikov} kernel for
applied work, consistent with the time-series finding of
\citet{KolokotronesStockWalker2024} regarding the optimality of Bartlett
among first-order kernels.

\subsection{Visual Summary}\label{visual-summary}

\begin{figure}

\centering{

\includegraphics{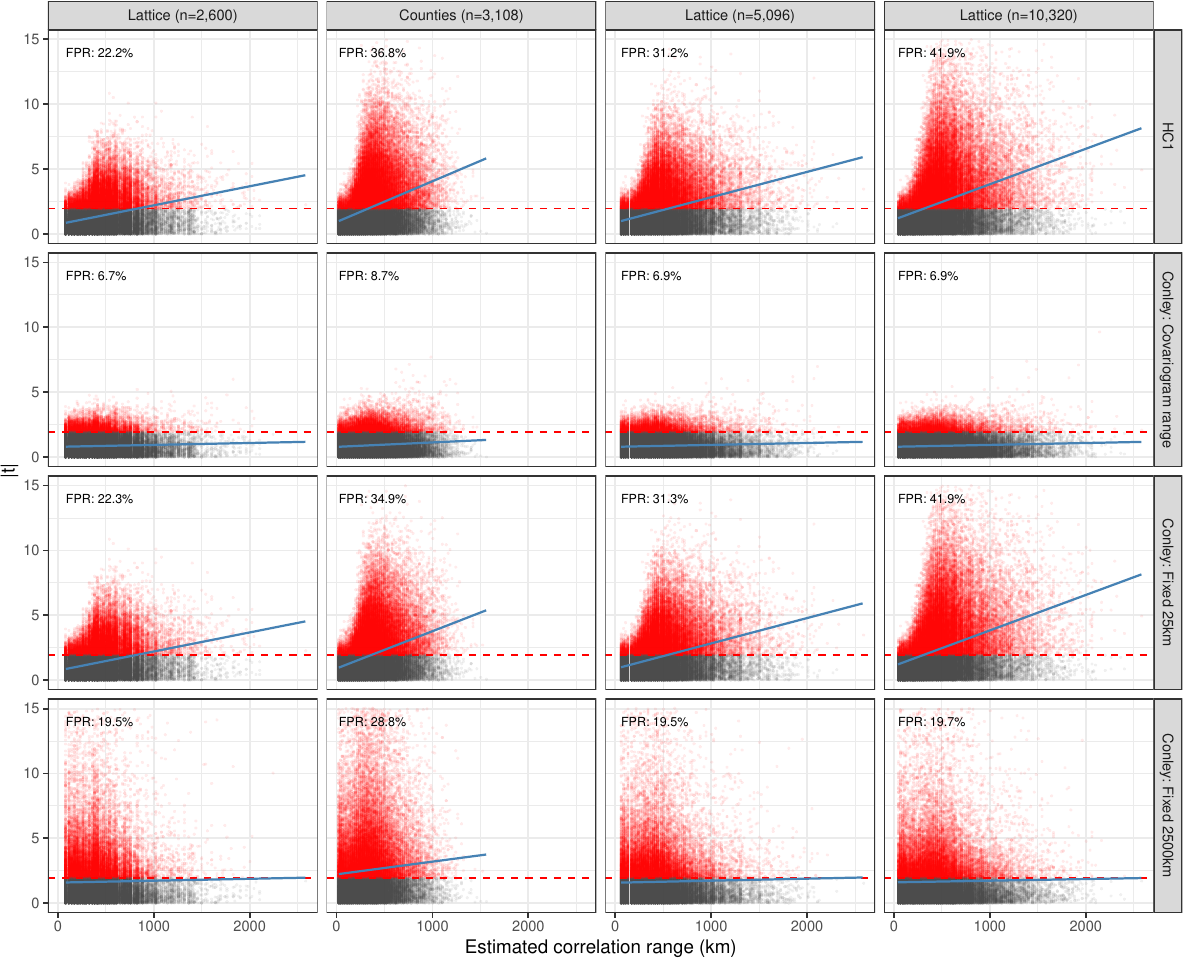}

}

\caption{\label{fig-scatters}Visual summary of the simulation results in
tables 1 and 2. Absolute \(t\)-statistics from 5,000 Monte Carlo
simulations plotted against the estimated correlation range. Each dot is
one simulation. Red dots indicate rejections (\(|t| > 1.96\)). The
annotation shows the false positive rate (FPR) across the entire sample.
Columns correspond to the four spatial samples; rows compare HC1,
spatial HAC errors using the proposed covariogram-range bandwidth
estimator (Epanechnikov kernel), a fixed narrow bandwidth (25 km), and a
fixed wide bandwidth (2,500 km). The blue line indicates a regression
line showing the decreasing performance of conventional errors as the
degree of spatial correlation increases. It can be visually inferred
that the proposed method provides reasonable size control even for
extreme degrees of spatial correlation}

\end{figure}

Figure~\ref{fig-scatters} provides a visual summary of the results
presented in Table~\ref{tbl-main} and Table~\ref{tbl-county}, plotting
every individual simulation run and illustrating how false rejection
rates increase with the degree of spatial correlation. Each column
corresponds to one of the four spatial samples; each row to a different
standard error type. The top row (HC1) shows that \(t\)-statistics grow
with the correlation range when both the outcome and regressor are
spatially correlated, producing the substantial overrejection documented
in the tables. The second row demonstrates the covariogram-range method:
the \(t\)-statistics remain centered below the critical value regardless
of the correlation range, with approximately 5\% of dots exceeding the
threshold. The third and fourth rows show the failure modes of fixed
bandwidths: a narrow bandwidth (25 km) behaves like HC1 when the true
range exceeds 25 km, and a wide bandwidth (2,500 km) deflates standard
errors and produces high rejection rates across the board.

\section{Empirical Application}\label{sec-empirical}

To illustrate the practical relevance of the proposed method, I apply it
to a regression using U.S. county-level data. The outcome variable is
the share of the Black population in each county, and the regressor is
the share of the Hispanic population---two variables that exhibit
well-known spatial clustering patterns across the United States. The
data is from the latest US census, accessed through IPUMS
\citep{IPUMS2025}.

\begin{table}

\caption{\label{tbl-empirical}Empirical application: Regression of Black
population share on Hispanic population share across 3,108 U.S.
counties. Standard errors in parentheses. Conley standard errors
computed using the Epanechnikov kernel.}

\centering{

[!h]
\centering\begingroup\fontsize{9}{11}\selectfont

\begin{tabular}[t]{lccc}
\toprule
Standard error type & Coefficient & SE & $t$-statistic\\
\midrule
HC1 & -0.1168 & 0.0101 & -11.51\\
Conley (25 km) & -0.1168 & 0.0105 & -11.10\\
\textbf{Conley (980 km, $\hat{\varsigma}$)} & \textbf{-0.1168} & \textbf{0.0884} & \textbf{-1.32}\\
Conley (500 km) & -0.1168 & 0.0766 & -1.53\\
Conley (2500 km) & -0.1168 & 0.0561 & -2.08\\
\bottomrule
\end{tabular}
\endgroup{}

}

\end{table}

Table~\ref{tbl-empirical} presents the results. The HC1 standard error
is the smallest, producing the largest \(t\)-statistic. The Conley
standard errors vary substantially with the bandwidth choice. At the
estimated covariogram range of approximately 980 km, the standard error
is the largest among all choices shown, consistent with the inverse-U
pattern. This is the appropriate bandwidth for this regression: the
estimated range reflects the spatial extent of residual correlation in
these demographic variables. At bandwidths of 500 km and 2,500
km---which might seem ``safer'' to an applied researcher---the standard
errors are actually smaller, illustrating precisely the misleading
effect documented in Section~\ref{sec-inverseu}.

This example demonstrates a scenario commonly encountered in applied
spatial research: widely used variables with strong spatial clustering
patterns that extend over large distances. The covariogram range method
correctly identifies this extent and produces appropriately sized
standard errors.

\section{Conclusion}\label{sec-conclusion}

This paper addresses the unresolved problem of bandwidth selection for
spatial HAC standard errors. I make three main contributions.

First, I document that the relationship between the kernel bandwidth and
the magnitude of spatial HAC standard errors follows an inverse-U shape.
This implies that both too narrow \emph{and} too wide bandwidths lead to
underestimated standard errors. The finding contradicts the common
advice that wider bandwidths are more conservative and establishes that
the bandwidth choice is not merely a matter of being ``generous'' with
the cutoff distance.

Second, I propose a simple, non-parametric bandwidth selector based on
the empirical covariogram of regression residuals. The estimator
identifies the distance at which residual covariation first crosses
zero---the correlation range---and uses this as the bandwidth for the
spatial HAC estimator. Under standard regularity conditions, the
estimator is consistent for the true correlation range.

Third, I show through extensive Monte Carlo simulations that the
proposed method controls the false positive rate at or near the nominal
5\% level across a wide range of spatial autocorrelation intensities and
sample configurations. A systematic comparison of six kernel functions
finds that the Bartlett and Epanechnikov kernels deliver the best size
control, extending the time-series finding of
\citet{KolokotronesStockWalker2024} to the spatial domain. At extreme
levels of spatial correlation, where the correlation range approaches a
substantial fraction of the spatial domain, the rejection rate rises to
approximately 8--10\%. This residual overrejection should be understood
as a fundamental limitation of spatial inference under very strong
dependence rather than a failure of the proposed selector: when
effective degrees of freedom become small, no HAC-type
estimator---whether based on fixed bandwidths, cluster-robust methods,
or data-driven selection---performs well. The key result is comparative:
across all configurations considered, the covariogram-range method
substantially improves size control relative to HC1, fixed narrow
bandwidths, and fixed wide bandwidths.

The proposed method is implemented in the R package
\texttt{SpatialInference}.

\subsection{Declaration of generative AI and AI-assisted technologies in
the manuscript preparation
process}\label{declaration-of-generative-ai-and-ai-assisted-technologies-in-the-manuscript-preparation-process}

During the preparation of this work the author used Claude (Anthropic)
in order to assist with manuscript drafting. After using this tool, the
author reviewed and edited the content as needed and takes full
responsibility for the content of the published article.

\section{References}\label{references}

\renewcommand{\bibsection}{}
\bibliography{MyLibrary}

\newpage

\section{Appendix}\label{sec-appendix}

\subsection{Details on Spatial HAC Estimation}\label{sec-appendix-hac}

The spatial HAC estimator in Equation~\ref{eq-vcov-conley} can be
written more explicitly. Let \(\hat{\varepsilon}_i\) denote the OLS
residual for observation \(i\), and let \(\mathbf{X}\) be the
\(n \times p\) regressor matrix. The estimator of the middle term is \[
\hat{\mathbf{\Omega}} = \sum_{i=1}^n \sum_{j=1}^n K\!\left(\frac{d_{ij}}{\varsigma}\right) \hat{\varepsilon}_i \hat{\varepsilon}_j \, \mathbf{x}_i \mathbf{x}_j',
\] where the double sum runs over all \(n^2\) pairs. In practice, this
is computed efficiently by exploiting the sparsity of the kernel: only
pairs with \(d_{ij} < \varsigma\) contribute non-zero terms. The
distance matrix is computed using the Haversine formula when working
with geographic (longitude, latitude) coordinates, and Euclidean
distance when working with projected coordinates.

The spatial HAC estimator nests several familiar estimators as special
cases:

\begin{itemize}
\tightlist
\item
  When \(\varsigma = 0\): \(\hat{V}_{\text{SHAC}}\) reduces to the HC
  estimator \citep{Eicker1963, Huber1967, White1980}.
\item
  When \(K(\cdot) = \mathbf{1}(\cdot)\) (Uniform kernel) and
  \(\varsigma\) defines group membership: \(\hat{V}_{\text{SHAC}}\)
  reduces to the cluster-robust estimator
  \citep{LiangZeger1986, Moulton1986}.
\end{itemize}

See \citet{KelejianPrucha2007} for a detailed treatment of the
asymptotic properties and \citet{BesterConleyHansen2011} for the
connection to cluster covariance estimators.

\subsection{Spatial Kernel
Visualizations}\label{spatial-kernel-visualizations}

\begin{figure}

\begin{minipage}{0.33\linewidth}

\centering{

\includegraphics{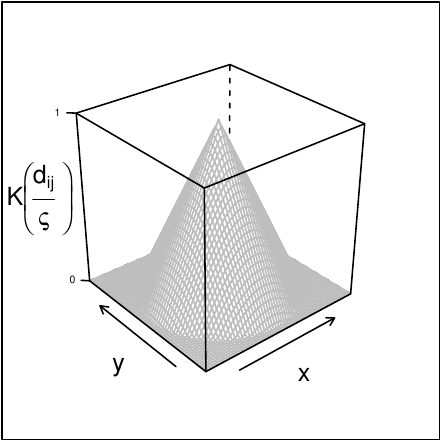}

}

\subcaption{\label{fig-kernels-1}Bartlett (triangular)}

\end{minipage}
\begin{minipage}{0.33\linewidth}

\centering{

\includegraphics{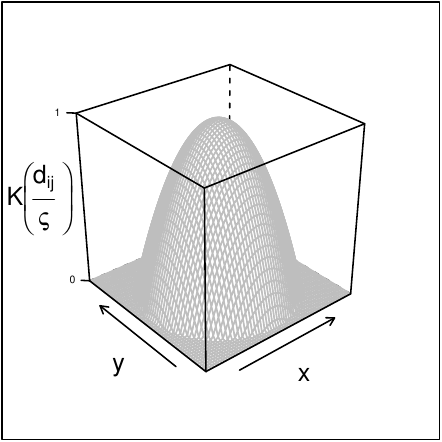}

}

\subcaption{\label{fig-kernels-2}Epanechnikov}

\end{minipage}
\begin{minipage}{0.33\linewidth}

\centering{

\includegraphics{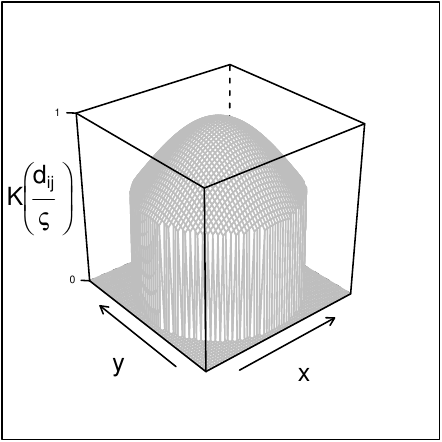}

}

\subcaption{\label{fig-kernels-3}Gaussian (truncated)}

\end{minipage}
\newline
\begin{minipage}{0.33\linewidth}

\centering{

\includegraphics{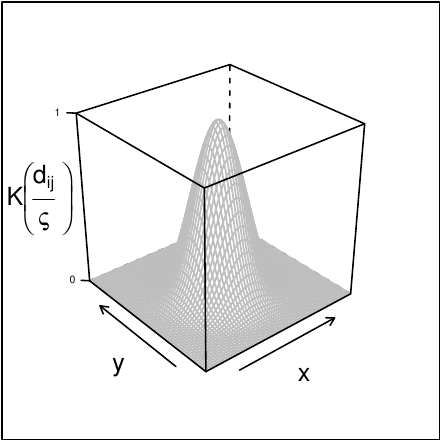}

}

\subcaption{\label{fig-kernels-4}Parzen}

\end{minipage}
\begin{minipage}{0.33\linewidth}

\centering{

\includegraphics{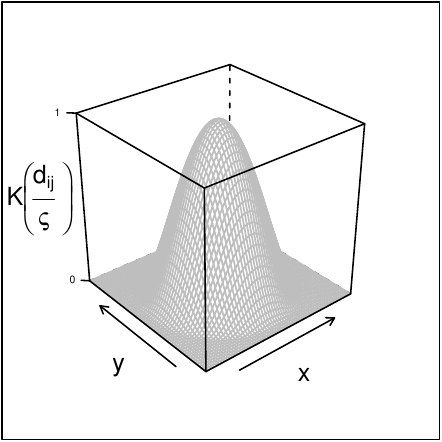}

}

\subcaption{\label{fig-kernels-5}Quartic Biweight}

\end{minipage}
\begin{minipage}{0.33\linewidth}

\centering{

\includegraphics{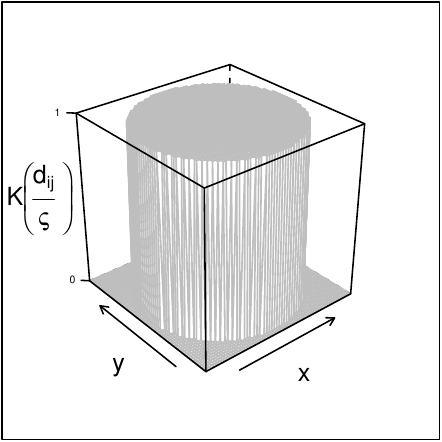}

}

\subcaption{\label{fig-kernels-6}Uniform (rectangular)}

\end{minipage}

\caption{\label{fig-kernels}All six two-dimensional kernels used in this
study, evaluated on a regular grid. Each panel shows the kernel weight
as a function of position relative to the center point.}

\end{figure}

\subsection{Generating Spatially Correlated Random
Fields}\label{sec-appendix-kriging}

Assume
\(\mathbf{z}(\mathbf{s}) = (z(\mathbf{s}_1), \ldots, z(\mathbf{s}_n))'\)
are observations with known mean at spatial locations
\(\mathbf{s}_1, \ldots, \mathbf{s}_n\). The random process can be
written in terms of the classical geostatistical model
\begin{equation}\phantomsection\label{eq-spatialprocess}{
\mathbf{z}(\mathbf{s}) = \boldsymbol{\mu}(\mathbf{s}) + \mathbf{e}(\mathbf{s}),
}\end{equation} where \(\mathbf{e}(\cdot)\) is a zero-mean random error
process with a spatially dependent component modeled by one of the
semivariogram functional forms introduced below, so that
\(E[\mathbf{z}(\mathbf{s})] = \boldsymbol{\mu}(\mathbf{s})\). For simple
kriging, the mean is set to zero.

The goal is to predict the value of \(z\) at a new location
\(\mathbf{s}_0\), i.e., to obtain an estimate \(\hat{z}(\mathbf{s}_0)\).
In ordinary kriging, the predicted value is a weighted linear
combination of the observed values:
\begin{equation}\phantomsection\label{eq-krigingprediction}{
\hat{z}(\mathbf{s}_0) = \sum_{i=1}^n \lambda_i \, z(\mathbf{s}_i).
}\end{equation} The weights \(\lambda_i\) are obtained by minimizing the
prediction error variance,
\(\text{Var}[\hat{z}(\mathbf{s}_0) - z(\mathbf{s}_0)]\). It can be shown
that solving the following block matrix system yields the optimal
weights \(\lambda_1, \ldots, \lambda_n\) \citep{Cressie1993}:
\begin{equation}\phantomsection\label{eq-kriging-system}{
\begin{bmatrix}
\boldsymbol{\Gamma} & \mathbf{1} \\
\mathbf{1}' & 0
\end{bmatrix}
\begin{bmatrix}
\boldsymbol{\lambda} \\
m
\end{bmatrix}
=
\begin{bmatrix}
\boldsymbol{\gamma}_0 \\
1
\end{bmatrix},
}\end{equation} where \(\boldsymbol{\Gamma}\) is the \(n \times n\)
matrix with entries
\([\boldsymbol{\Gamma}]_{ij} = \gamma(d(\mathbf{s}_i, \mathbf{s}_j))\),
and \(\boldsymbol{\gamma}_0\) is the \(n \times 1\) vector with entries
\([\boldsymbol{\gamma}_0]_i = \gamma(d(\mathbf{s}_i, \mathbf{s}_0))\).
The vector of weights is
\(\boldsymbol{\lambda} = (\lambda_1, \lambda_2, \ldots, \lambda_n)'\)
and \(m\) is a Lagrange multiplier ensuring
\(\sum_{i=1}^n \lambda_i = 1\). The vectors \(\mathbf{1}\) are of length
\(n\).

The defining element of this spatial process is the semivariogram
\(\gamma(\cdot)\). It determines the semivariance between different
pairs of observations that are distance
\(d(\mathbf{s}_i, \mathbf{s}_j)\) apart:
\begin{equation}\phantomsection\label{eq-semivariogram}{
\gamma(d(\mathbf{s}_i, \mathbf{s}_j)) = \frac{1}{2} \text{Var}[z(\mathbf{s}_i) - z(\mathbf{s}_j)].
}\end{equation} Note that the relation to the covariogram is
\(C(\mathbf{h}) = C(\mathbf{0}) - \gamma(\mathbf{h})\), where
\(C(\mathbf{0})\) denotes the variation at distance zero---referred to
as the ``nugget effect'' in the geostatistical literature. The
covariance between points becomes effectively zero at a distance \(h\)
equal to the range.

The shape of the random spatial process is determined by choosing one of
the various valid semivariogram model distance decay functions. For the
evaluation of standard error performance, only variations in the range
parameter \(\theta_2\) matter while everything else is held constant.
For completeness, the most commonly used functional forms in
geostatistics are:

\begin{longtable}[]{@{}
  >{\raggedright\arraybackslash}p{(\columnwidth - 2\tabcolsep) * \real{0.5000}}
  >{\raggedright\arraybackslash}p{(\columnwidth - 2\tabcolsep) * \real{0.5000}}@{}}
\caption{Semivariogram models used in geostatistical simulation. In all
specifications, \(\theta_1\) is the partial sill and \(\theta_2\) is the
range parameter.}\label{tbl-semivariograms}\tabularnewline
\toprule\noalign{}
\begin{minipage}[b]{\linewidth}\raggedright
Model
\end{minipage} & \begin{minipage}[b]{\linewidth}\raggedright
Semivariogram Specification
\end{minipage} \\
\midrule\noalign{}
\endfirsthead
\toprule\noalign{}
\begin{minipage}[b]{\linewidth}\raggedright
Model
\end{minipage} & \begin{minipage}[b]{\linewidth}\raggedright
Semivariogram Specification
\end{minipage} \\
\midrule\noalign{}
\endhead
\bottomrule\noalign{}
\endlastfoot
\textbf{Exponential} &
\(\gamma(h; \boldsymbol{\theta}) = \theta_1 \{ 1 - \exp(-h/\theta_2) \}\) \\
\textbf{Gaussian} &
\(\gamma(h; \boldsymbol{\theta}) = \theta_1 \{ 1 - \exp(-h^2/\theta_2^2) \}\) \\
\textbf{Mat\textquotesingle\{e\}rn} &
\(\gamma(h; \boldsymbol{\theta}) = \theta_1 \left( 1 - \frac{(h/\theta_2)^\nu K_\nu(h/\theta_2)}{2^{\nu-1} \Gamma(\nu)} \right)\),
where \(K_\nu(\cdot)\) is the modified Bessel function of the second
kind of order \(\nu\) and \(\Gamma(\cdot)\) is the Gamma function \\
\textbf{Spherical} &
\(\gamma(h; \boldsymbol{\theta}) = \begin{cases} \theta_1 \left( \frac{3h}{2\theta_2} - \frac{h^3}{2\theta_2^3} \right) & \text{for } 0 \leq h \leq \theta_2 \\ \theta_1 & \text{for } h > \theta_2 \end{cases}\) \\
\end{longtable}

The preferred choice in this paper is the Mat\textquotesingle\{e\}rn
semivariogram, with partial sill \(\theta_1 = 0.025\). Results are
qualitatively identical with Exponential and Spherical specifications.

In practice, simulation proceeds by: (i) defining a regular 10 km grid
covering the bounding box of the observation locations (with a 100 km
buffer); (ii) generating an unconditional realization of the random
field on this grid using \texttt{gstat}; (iii) extracting values at the
observation locations; and (iv) standardizing to zero mean and unit
variance. The range parameter \(\theta_2\) is varied to produce random
fields with increasing spatial autocorrelation. A disadvantage of this
simulation methodology is that the range cannot be steered with
precision and there is substantial variation across draws with the same
input values. However, the resulting random fields cover the entire
spectrum well and the method is thus well suited for the Monte Carlo
evaluation. The distribution of observations and the shape of the
spatial random fields are illustrated in Figure~\ref{fig-firstmaps}
below.

\subsection{Spatial Samples and Random Fields}\label{sec-appendix-maps}

\begin{figure}

\begin{minipage}{0.50\linewidth}

\centering{

\includegraphics{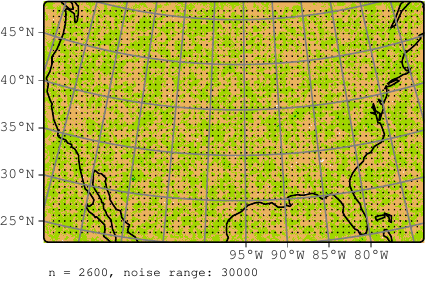}

}

\subcaption{\label{fig-firstmaps-1}Coarse lattice
(\(n \approx 2{,}600\)), range 30 km.}

\end{minipage}
\begin{minipage}{0.50\linewidth}

\centering{

\includegraphics{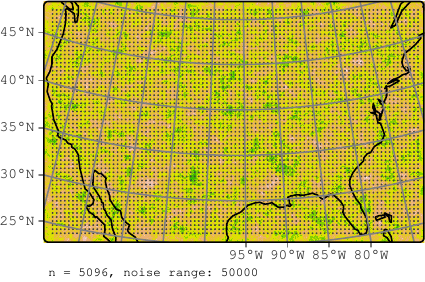}

}

\subcaption{\label{fig-firstmaps-2}Medium lattice
(\(n \approx 5{,}096\)), range 50 km.}

\end{minipage}
\newline
\begin{minipage}{0.50\linewidth}

\centering{

\includegraphics{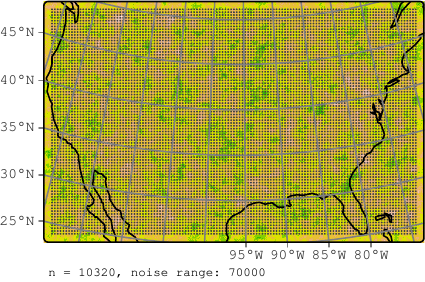}

}

\subcaption{\label{fig-firstmaps-3}Fine lattice
(\(n \approx 10{,}320\)), range 70 km.}

\end{minipage}
\begin{minipage}{0.50\linewidth}

\centering{

\includegraphics{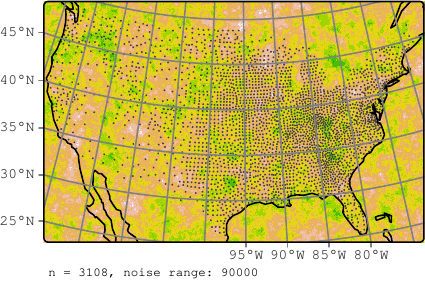}

}

\subcaption{\label{fig-firstmaps-4}County centroids (\(n = 3{,}108\)),
range 90 km.}

\end{minipage}

\caption{\label{fig-firstmaps}All four spatial point samples overlaid on
randomly generated spatial fields with increasing correlation ranges. In
the Monte Carlo experiments, each point is assigned the underlying value
of the spatial field at its location.}

\end{figure}

Figure~\ref{fig-firstmaps} illustrates the four spatial point samples
used throughout the study, each overlaid on a realization of a spatially
correlated random field with increasing range parameters. The coarsest
lattice (panel a) has large and equal spacing between points, while the
finest lattice (panel c) has little spacing. The county centroids (panel
d) exhibit highly unequal spacing, with dense coverage in the eastern
United States and sparser coverage in the west. As the range parameter
increases from 30 km to 90 km, the spatial fields become visibly
smoother, reflecting the growing extent of spatial dependence.

\subsection{Handling
Non-Stationarity}\label{sec-appendix-nonstationarity}

The covariogram range method relies on the assumption that the spatial
correlation structure is stationary (Assumption \ref{ass:stationarity}).
In practice, many spatial variables exhibit spatial trends---a form of
non-stationarity. Common examples include variables that vary
systematically with latitude (e.g., temperature-related outcomes) or
that cluster around geographic features (e.g., economic activity near
coasts or rivers).

Non-stationarity can be addressed within the proposed framework by
detrending before applying the covariogram method. The simplest approach
is to include location coordinates as control variables in the
regression: \[
y_i = \mathbf{x}_i' \boldsymbol{\beta} + f(\text{lon}_i, \text{lat}_i) + \varepsilon_i,
\] where \(f(\cdot)\) is a flexible function of geographic coordinates
(e.g., a polynomial in longitude and latitude). The covariogram is then
computed on the residuals from this augmented regression, which removes
the spatial trend and produces a residual process that is closer to
stationary.

The Monte Carlo experiments in Section~\ref{sec-montecarlo} include
specifications without longitude and latitude controls. The
covariogram-range method performs well in both cases, though including
location controls can improve size control when the dependent variable
exhibits strong spatial trends.

For variables with extreme non-stationarity---such as distance to a
fixed geographic point---the covariogram range method should be applied
to residuals from a detrended regression. In such cases, the false
positive rate after detrending is substantially reduced compared to the
raw regression.

\end{document}